%% file: main.tex
\documentclass[runningheads,orivec]{llncs}
\usepackage[T1]{fontenc}
\usepackage{graphicx}
\usepackage{times}
\usepackage{latexsym}
\usepackage{algorithm}
\usepackage{algpseudocode}
\usepackage{float}
\usepackage{amsmath}
\usepackage{esvect}
\usepackage{amssymb}     
\usepackage{pifont}      

\usepackage{multirow}
\usepackage{adjustbox}
\usepackage{tikz}
\usepackage{booktabs}
\usepackage[hidelinks]{hyperref}

\begin{document}
\title{More Efficient LLM Reranking with Whole-Pool, Setwise, Long-Context Language Models}
\titlerunning{More Efficient LLM Reranking}
%
\author{Hang Li\orcidID{0000-0002-5317-7227} \and
Chuting Yu\orcidID{0009-0008-5150-3843} \and \\
Bevan Koopman\orcidID{0000-0001-5577-3391} \and
Guido Zuccon\orcidID{0000-0003-0271-5563}}

\authorrunning{Hang Li et al.}

\institute{The University of Queensland, Queensland, Australia \\
\email{\{hang.li, v.yu, b.koopman, g.zuccon\}@uq.edu.au}}

\maketitle              
\begin{abstract}
\input{sections/abstract}
\end{abstract}
\input{sections/introduction}
\input{sections/related_work}
\input{sections/methodology}
\input{sections/experimental_setup}
\input{sections/results}
\input{sections/conclusion}

%
%
%
\bibliographystyle{splncs04}
\bibliography{references.bib}
\end{document}

%% file: sections/abstract.tex
LLM-based re-rankers produce rankings through repeated local comparisons (listwise, pairwise or pointwise), requiring many sequential model calls. We study how long-context LLMs can drastically reduce this computation when the entire retrieved candidate pool fits within the context window. We introduce Whole-Pool Setwise re-ranking, where each comparison ranks all the entire candidate pool, and propose DualEnd Setwise, which jointly selects the candidates predicted to be most and least relevant. By filling the ranking from both ends, DualEnd constructs a complete ranking of 100 candidates in 50 LLM comparisons. Experiments with nine open-weight LLMs on TREC DL19 and DL20 show that this requires 59.4\% fewer comparisons than previous top-oriented windowed Setwise with heapsort and 88.8\% fewer than top-oriented windowed Setwise with bubblesort, even though those baselines target only the top-10 rankings while DualEnd targets the full ranking. DualEnd's nDCG@100 is within 0.008 of the single-end whole-pool top-oriented approach, while approximately halving its token consumption and ranking time. Across six BEIR datasets, DualEnd reduces mean token consumption and ranking time by 49.4\% and 50.8\%, respectively, relative to single-end whole-pool top-oriented approach. These results demonstrate that DualEnd Setwise enables complete re-ranking with substantially fewer LLM comparisons and competitive effectiveness across several backbones. 
	Implementation, results, and prompt templates available at \url{https://github.com/hanglics/Whole-Pool-Setwise}.

\keywords{Large Language Models \and Setwise \and Reranking \and Efficiency}

%% file: sections/introduction.tex
\section{Introduction}
\label{sec:introduction}

Large language models (LLMs) are increasingly used as effective rerankers~\cite{zuccon2025r2llms}. Given $N$ candidate passages (the \emph{candidate pool}) returned by a first-stage retriever, re-ranking produces a top-$k$ output, where $k\le N$, using independent candidate scoring~\cite{sachan2022upr}, candidate comparisons~\cite{qin2024prp}, or direct generation of an ordering~\cite{sun2023rankgpt}. A complete re-ranking involves ranking all $N$ (i.e., $k=N$), making the rankings available at different depths $\le N$. This is particularly useful when the required cutoff varies or a downstream task needs access beyond a fixed top-$k\le N$ list. The computational challenge is to obtain that complete ranking without an excessive sequence of costly LLM calls.

Setwise re-ranking selects the most relevant candidate from a small subset of the entire candidate pool and combines these local decisions through a sorting procedure~\cite{zhuang2024setwise}. Repeated comparisons among small subsets can require substantial sequential LLM computations even for only producing a top-10 ranking; extending to deeper ranking outputs thus requires more comparisons and LLM computations. Long-context models offer a potential alternative: their larger context windows can accommodate the entire candidate pool when the rendered prompt fits~\cite{liu2025sliding}. However, even when the entire candidate pool fits the model context, there are multiple ways to produce the final ranking. A model may directly generate a complete ranking from the candidate pool, or make repeated selections over smaller subsets and progressively build the ranking. These strategies induce different trade-offs: including more candidates in each comparison produces longer inputs but typically requires fewer calls. Consequently, fewer model calls do not necessarily translate into lower token consumption or shorter elapsed time.

This paper introduces Whole-Pool Setwise re-ranking, which inputs the entire candidate pool in each comparison. Whole-pool selection replaces repeated comparisons over small subsets and the subsequent sorting procedures they require. We further propose Whole-Pool DualEnd (WP-DE), which jointly selects the predicted best and worst candidates to build the ranking from both ends of the final list, placing the best candidate at the top and the worst at the bottom. The selected candidates are then removed from the candidate pool, reducing the candidates pool size at each step. We also consider two single-end variants: 1) WP-T (Top) which builds the ranking from the top, only selecting the best candidate; and 2) WP-B (Bottom) which builds the ranking from bottom, only selecting the worst candidate; both still input the entire candidate pool. Comparing these variants with Whole-Pool DualEnd quantifies the effectiveness–computation trade-off introduced by DualEnd selection for complete ranking. All variants use existing instruction-tuned models without additional training.

We evaluate 9 open-weight LLMs on TREC DL19, DL20 and BEIR datasets. At pool size 100, WP-DE builds complete ranking in 50 comparisons, substantially fewer than all Setwise methods require when only outputting top-10. Mean top-10 effectiveness trails the top-oriented windowed Setwise baselines, with backbone-dependent gaps. Separate matched top-10 comparisons on three backbones confirm that Whole-Pool selection also reduces the comparison workload for that objective. For complete ranking, DualEnd reduces the evaluated single-end variants' 98 comparisons to only 50 comparison. Compared with WP-T, it approximately halves token consumption and ranking time, with nDCG@100 within 0.008 of WP-T for seven of nine models. BEIR experiments and controls for selection scope and candidate ordering examine how broadly these computational savings translate into competitive effectiveness.

%% file: sections/related_work.tex
\section{Related Works}
\label{sec:related_work}

LLM-based re-ranking methods differ in a number of ways: the candidates considered per inference, the form of their relevance judgments, and how to utilise these judgments to build the final ranking. Our Whole-Pool Setwise connects these choices through Whole-Pool visibility and DualEnd selection.

Unsupervised Passage Re-ranker obtains pointwise relevance signals from query likelihood conditioned on each candidate passage~\cite{sachan2022upr}, while RankSteer calibrates pointwise judgments through activation steering~\cite{wang2026ranksteer}. Pairwise Ranking Prompting uses explicit candidate comparisons to elicit relative preferences~\cite{qin2024prp}. The comparison set defines which alternatives the model evaluates together, connecting the form of a relevance judgment to its input scope. WP-Setwise builds on this comparative formulation, selecting from all remaining candidates rather than judging individual candidate or pairs.

Generative listwise methods jointly present passages and generate ordered passage identifiers. RankGPT and Ma et al. use overlapping windows~\cite{sun2023rankgpt,ma2023zeroshotlistwise}, while RankVicuna and RankZephyr distill LLM-generated rankings into open models~\cite{pradeep2023rankvicuna,pradeep2023rankzephyr}. FIRST instead derives an ordering from first-position candidate-identifier logits with ranking-aware training~\cite{reddy2024first}. WP-Setwise also limits output generation, but builds rankings through repeated endpoint selections.

The availability of long-context models make it possible for multiple candidate comparisons to be extended out to the entire candidate pool. FullRank~\cite{liu2025sliding} explores single-inference full ranking with complete-ranking labels and an importance-aware objective for supervised training. WP-Setwise still uses the entire candidate pool but replaces full-permutation generation with compact selections from a shrinking candidate pool, using existing instruction-tuned models without additional ranking-specific training. Single-pass ranking avoids repeated candidate presentation across calls, whereas DualEnd reduces the actual number of comparisons needed. These are distinct sources of computational savings.

The selection strategy in WP-Setwise extends the comparison-based formulation used by Setwise, which combines local best-candidate choices through a sorting algorithm such as heapsort or bubblesort~\cite{zhuang2024setwise}. Setwise Insertion reduces comparisons for top-$k$ construction and exploits initial order~\cite{podolak2025setwiseinsertion}, while TourRank aggregates advancement scores across grouped tournaments~\cite{chen2025tourrank}. Rank-R1 improves setwise relevance reasoning through reinforcement learning~\cite{zhuang2025rankr1}. WP-Setwise extends the selection formulation to the whole remaining pool, while DualEnd commits predicted best and worst candidates directly to opposite ends of the ranking.

How the candidate passages in the pool are input into an LLM is an important design consideration. Prior work considered order effects in long contexts~\cite{liu2024lostmiddle}, label bias in multiple-choice selection~\cite{zheng2024mcqbias}, and position bias in LLM judging~\cite{shi2025positionbias}. For listwise ranking, permutation self-consistency aggregates rankings from permuted inputs~\cite{tang2024foundmiddle}, while RankZephyr reports robustness to initial document ordering~\cite{pradeep2023rankzephyr}. Learning from all these prior works informed how we do our experiments on candidate ordering, candidate labels, and other diagnostics in Section~\ref{sec:presentation-sensitivity}.

%% file: sections/methodology.tex
\section{Whole-Pool Setwise Re-ranking}
\label{sec:methodology}

For a query $q$, let $D=(d_1,\ldots,d_N)$ be the candidate pool of $N$ candidates (i.e., passages) from first-stage retrieval. A Windowed Setwise (SW) reranker compares a series of small overlapping subsets from the entire candidate pool $D$ each round until full coverage, then combines these local decisions through a sorting procedure to produce the final ranking~\cite{zhuang2024setwise}. Conversely, Whole-Pool Setwise (WP-Setwise) instead presents all of $D$ in each comparison, and the sorting procedure afterwards is no longer needed, as each comparison gives the current global best/worst candidate from the entire candidate pool. Figure~\ref{fig:setwise-comparison} gives illustrative examples of the Windowed Setwise and Whole-Pool Setwise methods along with their abbreviations.

\input{figures/setwise_comparison}

Whole-Pool Setwise-Top (WP-T) builds the ranking from the top by selecting the best candidate from candidate pool at each round; Whole-Pool Setwise-Bottom (WP-B) builds the ranking from the bottom by selecting the worst candidate from candidate pool at each round. Whole-Pool Setwise-DualEnd (WP-DE) combines the two, selecting both the best and worst candidates in one call and building the ranking from top and bottom together. Algorithm~\ref{alg:dualend} defines the WP-DE method: In each round, \textsc{DualEndCompare} receives the query $q$ and candidate pool $D$, and requests \texttt{Best: i, Worst: j}, where distinct labels $i,j\in\{1,\ldots,|D|\}$. The best and worst candidates $(d^{+},d^{-})$ are then inserted into the top and bottom positions of the final ranking; these two are then removed from the candidate pool. These swaps can change the remaining pool's presentation order. For a complete ranking with odd $N$, the last remaining candidate in the candidate pool occupies the middle position without extra comparison.

\begin{algorithm}[t]
\caption{Whole-Pool Setwise-DualEnd}
\label{alg:dualend}
\footnotesize
\begin{algorithmic}[1]
\Require Query $q$; ordered candidates $D=(d_1,\ldots,d_N)$
\Ensure Full output ranking $R$
\State $R\gets D$; $h\gets1$; $t\gets N$
\While{$h<t$}
    \State $D\gets R[h:t]$
    \State Label candidates in $D$ as $1,\ldots,|D|$
    \State $(d^{+},d^{-})\gets\Call{DualEndCompare}{q,D}$
    \State Swap $d^{+}$ with $R[h]$
    \State Locate $d^{-}$ in the updated $R$
    \State Swap $d^{-}$ with $R[t]$
    \State $h\gets h+1$; $t\gets t-1$
\EndWhile
\State \Return $R$
\end{algorithmic}
\end{algorithm}

WP-DE uses $\lfloor N/2\rfloor$ LLM comparisons to build a complete ranking. Intuitively, selecting one candidate until one remains would require $N-1$ comparisons, our evaluated WP-T/WP-B implementations instead order the final pair by first-stage scores, breaking ties by original order, and use $N-2$ comparisons for $N\ge2$.\footnote{The final candidates pair are often highly similar and difficult to distinguish. Therefore, we avoid additional LLM calls and order them using their first-stage retrieval scores.} At $N=100$, WP-DE therefore uses 50 comparisons versus 98 for either single-end variant.

WholePool methods do not have a specific candidate pool size limit beyond LLM context input limits. With each candidate length bounded by $L$ tokens, complete ranking requires $O(N)$ comparisons but $O(LN^2)$ aggregate candidate-token input because each round presents the shrinking pool. Context, memory, and runtime therefore constrain practical pool sizes.

When doing top-$k<N$ ranking, WP-DE stops after fixing $k$ top positions or completing the ranking, using $\min(k,\lfloor N/2\rfloor)$ comparisons and returning the requested top-$k$ ranking. Each comparison still selects both endpoints. At $N=100$ and $k=10$, WP-T and WP-DE both use 10 comparisons.

%% file: figures/setwise_comparison.tex
\begin{figure}[t]
\centering
\includegraphics[width=\textwidth]{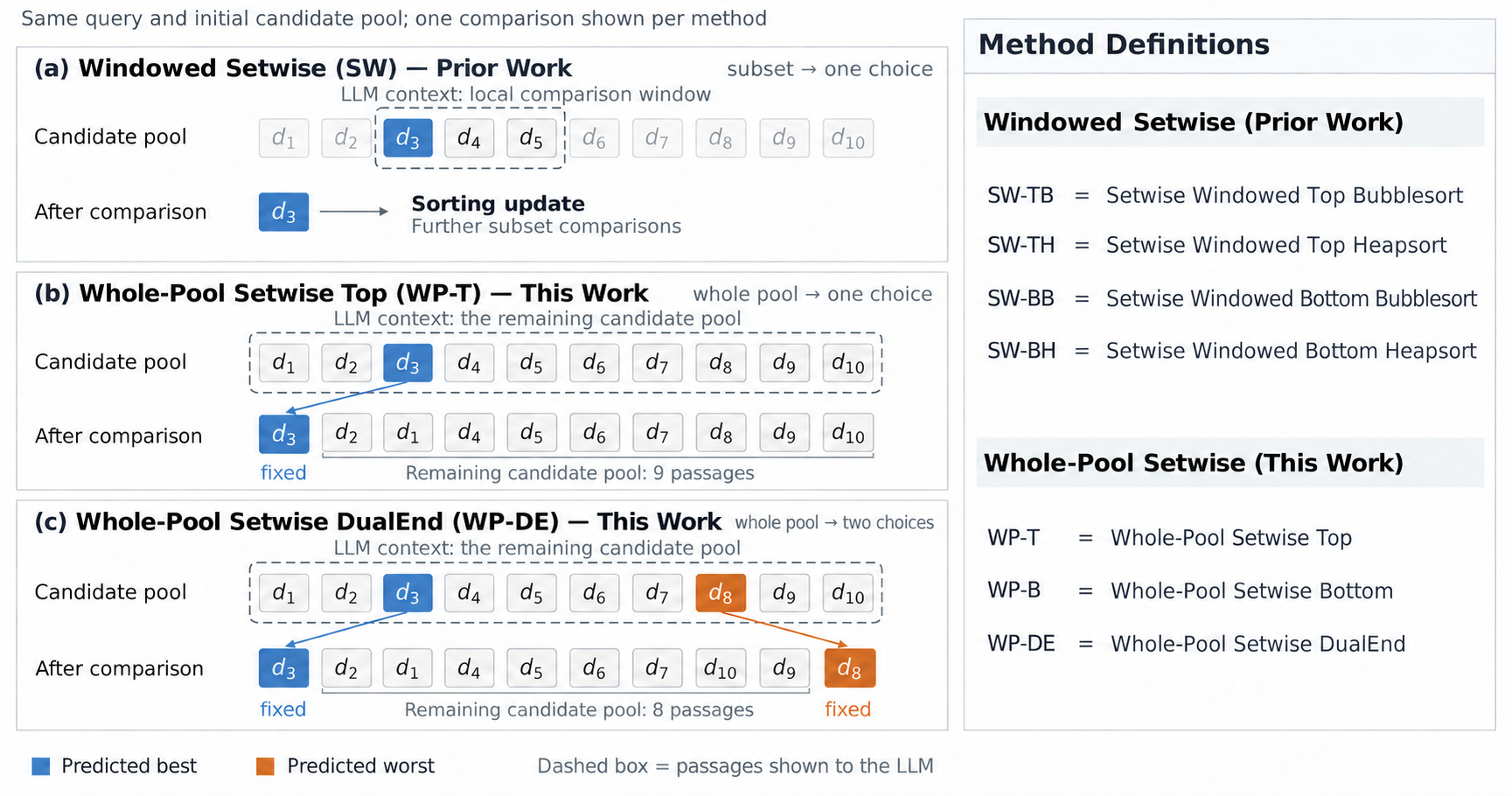}
\vspace{-20pt}
\caption{One LLM comparison in (a) Windowed Setwise, (b) WP-T and (c) WP-DE. Dashed boxes mark input candidates, blue and orange indicate predicted best and worst candidates. Windowed selections require sorting, Whole-Pool selections fix one or both endpoints through swaps.}
\label{fig:setwise-comparison}
\vspace{-15pt}
\end{figure}

%% file: sections/experimental_setup.tex
\vspace{-11pt}
\section{Experimental Setup}
\label{sec:experimental-setup}
\vspace{-6pt}
We evaluate on TREC DL19~\cite{craswell2020dl19} and DL20~\cite{craswell2021dl20} (43 and 54 judged queries), using MS MARCO Passage V1~\cite{bajaj2016msmarco}. BM25 top-$N$ initial candidate pools are taken from Pyserini~\cite{lin2021pyserini}. We test $N\in\{10,20,30,40,50,100\}$, with $N=100$ for main comparisons.

Instruction-tuned backbones used are Qwen3.5 0.8B, 2B, 4B, 9B and 27B (Q0.8B--Q27B)~\cite{qwen2026qwen35}, Llama3.1-8B (L8B)~\cite{dubey2024llama3}, and Ministral3 3B, 8B and 14B (M3B--M14B)~\cite{liu2026ministral3}. Setwise baselines~\cite{zhuang2024setwise} combine top- or bottom-oriented selection with bubblesort or heapsort (SW-TB, SW-TH, SW-BB, SW-BH), using three-candidate windows and alphabetical labels. Top-oriented variants stop after ten selections (top-$k=10$), bottom-oriented variants need to sort the entire pool before returning the top-10 ranking. WP-T, WP-B and WP-DE use numeric labels\footnote{Windowed Setwise takes a small subset from the candidate pool, Whole-Pool Setwise instead takes the entire candidate pool, therefore alphabetical labels are insufficient for WP methods.} and produce complete rankings.

By default, Whole-Pool Setwise parsing tolerates format variations from the models' generations, we allow up to three retries for failures, and use deterministic recovery based on original candidate pool rankings and scores from BM25. All experiments use 1 NVIDIA H100 GPU, 4 CPU cores and 512\,GB memory.

We report nDCG@10 and nDCG@$N$, using $N=100$ for fixed candidate pool complete rankings. All results are computed using averages to reflect mean effectiveness: DL results are averaged across datasets and models, and pool-size results are additionally averaged across the six pool sizes. Per-query comparisons, prompt-plus-completion token counts, and ranking time are averaged in the same way\footnote{Comparisons exclude additional retry generations, token totals retain the final generation per comparison, ranking time excludes the measured duration of additional retries. Table~\ref{tab:matched-output-baselines} instead includes all generation attempts, their tokens and measured query wall time.}. Statistical tests and correction schemes are described in corresponding table captions.

We benchmark against two methods using Q9B, L8B and M8B backbones at $N=100$: TourRank-2~\cite{chen2025tourrank} and FullRank~\cite{liu2025sliding} in Section~\ref{sec:external-baselines}. Within each comparison, methods have the same requested output: the top 10 candidates on DL19/20 or a complete ranking of all 100 candidates on DL19. Matched SW variants reuse the initial BM25 ranking, TourRank-2 uses two tournaments with the publicly available codebase. Our model-matched zero-shot adaptation of FullRank uses a 100-word candidate length limit and the same configurations from the original work.

We further evaluate on BEIR~\cite{thakur2021beir}, covering DBpedia, FiQA, NFCorpus, SciFact, Touche-2020 and TREC-COVID. WP-T and WP-DE use all nine backbones without additional training, they both fully re-rank identical BM25 candidate pools capped at 100. Shorter pools are retained, with identical evaluated queries\footnote{NFCorpus evaluation includes 308 queries, fifteen additional judged queries have no returned BM25 candidates.}. nDCG@10, nDCG@100 results are averaged as in the main experiments to obtain mean effectiveness.

Additional experiments examine bottom updates, window size, candidate ordering, candidate labels, ranking combination, and repeatability using Q9B, L8B and M8B at $N=100$; dataset scopes and individual conditions are specified alongside the results.

%% file: sections/results.tex
\section{Results}
\label{sec:results}


\subsection{Effectiveness and Computational Cost}
\label{sec:main-effectiveness-cost}

All 3 Whole-Pool methods build complete rankings using fewer LLM comparisons   (Comp./q) than the 4 windowed methods (SW-*) require for only top-10 output. Figure~\ref{fig:calls-vs-ndcg10} provides a visual overview. WP-DE produces a full ranking with 50 comparisons, 59\%--98\% fewer than the SW variants. This difference reflects how the windowed methods produce their outputs: SW-TB and SW-TH terminate once ten top-ranked candidates have been selected, whereas the bottom-oriented variants must resolve the remainder of the pool before the top-10 ranking is determined. Moreover, producing deeper rankings with the top-oriented windowed methods would require additional comparisons, while WP-DE’s 50-comparison budget already resolves all 100 candidates.

\input{figures/calls_vs_ndcg100}

\begin{table}[t]
\centering
\small
\setlength{\tabcolsep}{3pt}
\caption{Effectiveness and computational cost when candidate-pool $N=100$ on DL19/20. SW methods return top-10 output, WP methods rank all 100. In Panels A/B, \textbf{bold} marks column maxima and $\uparrow$/$\downarrow$ mark significant differences above/below WP-DE (paired t-tests, Bonferroni correction, $\alpha=0.05$). Comp./q counts LLM comparisons, Tok./q reports prompt-plus-completion tokens in thousands, and s/q reports ranking time in seconds.}
\label{tab:dl-pool100-main}
\begin{adjustbox}{max width=\textwidth}
\begin{tabular*}{\textwidth}{@{\extracolsep{\fill}}lrrrrrrrrrr@{}}
\toprule
Method & Q0.8B & Q2B & Q4B & Q9B & Q27B & L8B & M3B & M8B & M14B & Avg. \\
\midrule
\multicolumn{11}{l}{\textit{Panel A: Top-10 Effectiveness --- nDCG@10}} \\
\midrule
BM25 & 0.493$^{\uparrow}$ & 0.493 & 0.493$^{\downarrow}$ & 0.493$^{\downarrow}$ & 0.493$^{\downarrow}$ & 0.493$^{\downarrow}$ & 0.493$^{\downarrow}$ & 0.493$^{\downarrow}$ & 0.493$^{\downarrow}$ & 0.493 \\
SW-TB & \textbf{0.545$^{\uparrow}$} & 0.627$^{\uparrow}$ & 0.698$^{\uparrow}$ & 0.713 & \textbf{0.733} & \textbf{0.663$^{\uparrow}$} & \textbf{0.671} & 0.703 & 0.708 & \textbf{0.673} \\
SW-TH & 0.514$^{\uparrow}$ & \textbf{0.630$^{\uparrow}$} & \textbf{0.702$^{\uparrow}$} & \textbf{0.717} & 0.730 & 0.637 & 0.663 & \textbf{0.705} & 0.708 & 0.667 \\
SW-BB & 0.161$^{\downarrow}$ & 0.569$^{\uparrow}$ & 0.606 & 0.656 & 0.716 & 0.556$^{\downarrow}$ & 0.632 & 0.644$^{\downarrow}$ & 0.641$^{\downarrow}$ & 0.576 \\
SW-BH & 0.203$^{\downarrow}$ & 0.513 & 0.555$^{\downarrow}$ & 0.632$^{\downarrow}$ & 0.714 & 0.522$^{\downarrow}$ & 0.605 & 0.583$^{\downarrow}$ & 0.591$^{\downarrow}$ & 0.546 \\
WP-T & 0.503$^{\uparrow}$ & 0.549$^{\uparrow}$ & 0.662 & 0.706 & 0.719 & 0.622 & 0.630 & 0.700 & 0.713 & 0.645 \\
WP-B & 0.446 & 0.482 & 0.518$^{\downarrow}$ & 0.623 & 0.676 & 0.573 & 0.548$^{\downarrow}$ & 0.634$^{\downarrow}$ & 0.613$^{\downarrow}$ & 0.568 \\
WP-DE & 0.401 & 0.502 & 0.658 & 0.692 & 0.719 & 0.609 & 0.649 & 0.700 & \textbf{0.716} & 0.627 \\
\midrule
\multicolumn{11}{l}{\textit{Panel B: Complete-Ranking Effectiveness --- nDCG@100 (No sorting algorithm required)}} \\
\midrule
WP-T & \textbf{0.498}$^{\uparrow}$ & \textbf{0.517} & \textbf{0.560} & \textbf{0.577} & \textbf{0.579} & \textbf{0.547} & 0.548 & \textbf{0.572} & \textbf{0.575} & \textbf{0.552} \\
WP-B & 0.460 & 0.479 & 0.491$^{\downarrow}$ & 0.539$^{\downarrow}$ & 0.562 & 0.517 & 0.502$^{\downarrow}$ & 0.545 & 0.538$^{\downarrow}$ & 0.515 \\
WP-DE & 0.435 & 0.494 & 0.557 & 0.572 & 0.578 & 0.540 & \textbf{0.549} & 0.566 & \textbf{0.575} & 0.541 \\
\bottomrule
\end{tabular*}
\end{adjustbox}
\smallskip
\begin{tabular*}{\textwidth}{@{\extracolsep{\fill}}lrrrrrrr@{}}
\toprule
\multicolumn{8}{@{}l}{\textit{Panel C: Mean per-Query Computational Cost}} \\
\midrule
Measure & SW-TB & SW-TH & SW-BB & SW-BH & WP-T & WP-B & WP-DE \\
\midrule
Output & Top 10 & Top 10 & Top 10 & Top 10 & Full 100 & Full 100 & Full 100 \\
Comp./q & 447.7 & 123.1 & 2500.0 & 436.3 & 98.0 & 98.0 & 50.0 \\
Tok./q (K) & 249.8 & 68.5 & 1380.7 & 244.7 & 506.7 & 501.1 & 254.7 \\
Sec./q & 87.8 & 24.2 & 388.8 & 67.8 & 90.8 & 67.3 & 42.0 \\
\bottomrule
\end{tabular*}
\vspace{-15pt}
\end{table}

Table~\ref{tab:dl-pool100-main} gives more detailed results. 
Panel A shows top-10 effectiveness trade-off. On average, WP-DE trails SW-TB and SW-TH by 0.046 and 0.040 nDCG@10. These differences narrow substantially with larger backbones: on Q27B and M8B, WP-DE is within 0.014 of both baselines, while on M14B it exceeds both by 0.008. Although the top-oriented SW variants perform significantly better on several smaller backbones, the differences are not significant on the larger models. WP-DE also outperforms both bottom-oriented windowed variants on eight of nine backbones.

Because each Whole-Pool comparison processes all candidates remaining in the pool, fewer comparisons do not necessarily translate into lower token consumption or query latency. In Table~\ref{tab:dl-pool100-main}, Panel C, WP-DE is faster than SW-TB, SW-BB, and SW-BH, while consuming similar tokens to SW-TB and SW-BH and substantially fewer than SW-BB. SW-TH is more efficient when only a top-10 ranking is required. The comparison, however, differs in output depth: SW-TH returns only the top 10, whereas WP-DE produces a complete 100 ranking.

For complete rankings (Panel B of Table~\ref{tab:dl-pool100-main}), WP-DE reduces comparisons from 98 to 50 relative to WP-T and WP-B, approximately halving token consumption and reducing ranking time. Its nDCG@100 remains within 0.008 of WP-T on seven of nine backbones, with smaller backbones showing larger drops. Mean nDCG@10 is 0.018 lower than WP-T, with significant differences on the small backbones. Thus, DuelEnd methods substantially reduce repeated pool processing while retaining similar full-ranking effectiveness to WP-T.

In the main $N=100$ runs, WP-DE records an average of 0.066 parser fallback events per query under the recovery policy described in Section~\ref{sec:experimental-setup}. The mean ranking time of 42 seconds per query makes the current implementation most relevant to offline or latency-tolerant workflows requiring complete candidate ranking. 

\begin{table}[t]
\centering
\small
\setlength{\tabcolsep}{3pt}
\caption{Top-10 and complete-ranking comparisons at $N=100$ using Q9B, L8B and M8B. Panel A averages DL19/20 within each backbone, then across backbones; Panel B averages the three backbones on DL19. WP top-10 runs stop after ten top selections. Comp./q counts all generation attempts; Tok./q includes prompt and completion tokens in thousands; s/q is ranking time in seconds.}
\label{tab:matched-output-baselines}
\begin{tabular*}{\textwidth}{@{\extracolsep{\fill}}lrrrrr@{}}
\toprule
Method & nDCG@10 & nDCG@100 & Comp./q & Tok./q (K) & s/q \\
\midrule
\multicolumn{6}{@{}l}{\textit{Panel A: Top-10 Output}} \\
SW-TB & 0.693 & -- & 455.8 & 253.9 & 75.3 \\
SW-TH & 0.686 & -- & 126.2 & 69.8 & 21.0 \\
TourRank-2 & 0.694 & -- & 26.07 & 43.0 & 22.2 \\
WP-T & 0.676 & -- & 10.0 & 92.2 & 14.0 \\
WP-DE & 0.666 & -- & 10.0 & 87.9 & 15.6 \\
\midrule
\multicolumn{6}{@{}l}{\textit{Panel B: Complete Ranking}} \\
FullRank (Zero-Shot) & 0.668 & 0.552 & 1.0 & 9.3 & 22.5 \\
WP-DE & 0.687 & 0.565 & 50.0 & 250.1 & 54.5 \\
\bottomrule
\end{tabular*}
\vspace{-10pt}
\end{table}

\subsection{Top-10 and Complete-Ranking Comparisons}
\label{sec:external-baselines}

The preceding comparison evaluates SW and WP methods with different output ranking depths. Table~\ref{tab:matched-output-baselines} compares methods with the same output depth: either top-10 (Panel A) or complete 100 (Panel B). For top-10, WP-T and WP-DE each require only 10 comparisons, representing 92.1\% fewer comparisons than SW-TH and 97.8\% fewer than SW-TB. Both methods fix one top position per comparison and therefore require the same number of comparisons to produce the top 10. WP-DE, however, also fixes a bottom position at each step, progressively shrinking the remaining candidate pool and consequently consuming fewer tokens than WP-T. This saving comes with a modest effectiveness and latency trade-off: WP-T achieves higher mean nDCG@10 and lower ranking time, while WP-DE trades some of this advantage for the additional bottom selections. Relative to the windowed baselines, WP-DE trails SW-TH and SW-TB by 0.020 and 0.027 mean nDCG@10, respectively, but both Whole-Pool variants remain faster, with token consumption between that of SW-TH and SW-TB. Although the bottom selections do not reduce the number of comparisons for top-10 output, they become more beneficial for complete 100 ranking, where resolving both ends simultaneously reduces the number of subsequent comparisons required.

Table~\ref{tab:matched-output-baselines} Panel A also compares the Whole-Pool methods with TourRank-2 for top-10 ranking output. WP-T and WP-DE each require ten comparisons, 61.6\% fewer than TourRank-2's 26.07. Despite using more comparisons, TourRank-2 achieves higher mean nDCG@10 while consuming fewer tokens, highlighting that LLM call count alone does not determine inference cost. By operating on smaller candidate groups from the entire pool, TourRank-2 avoids repeatedly presenting the full remaining pool, allowing a larger number of comparisons to incur lower overall token consumption.

For complete ranking on DL19 (Panel B), WP-DE outperforms the zero-shot FullRank adaptation in nDCG@100 on all three backbones, and also achieves higher mean nDCG@10 with model-dependent gains. The two methods have different computation–effectiveness trade-off: FullRank generates a complete permutation in a single call and therefore avoids repeated candidate presentations, whereas WP-DE builds the ranking iteratively by resolving both ends of the candidate pool across multiple comparisons. WP-DE therefore achieves higher effectiveness at the cost of additional comparisons and token consumption.

\subsection{Generalization Across Retrieval Datasets}
\label{sec:beir-generalization}

\begin{table}[t]
\centering
\small
\setlength{\tabcolsep}{3pt}
\caption{BEIR effectiveness and complete-ranking cost on identical BM25 candidate pools capped at 100. Superscripts $a$ and $b$ indicate significant differences from BM25 and WP-T with WP-DE respectively, within each dataset and metric (two-sided paired $t$-tests with Bonferroni correction, $\alpha=0.05$). WP-T and WP-DE effectiveness scores average nine backbones; Avg. averages the six datasets. Queries gives the evaluated query count. Costs are per-query means averaged equally over the 54 dataset--backbone configurations for each method. Tokens include prompt and completion tokens.}
\label{tab:beir-generalization}
\begin{tabular}{lrrrrrrr}
\toprule
& & \multicolumn{3}{c}{nDCG@10} & \multicolumn{3}{c}{nDCG@100} \\
\cmidrule(lr){3-5}\cmidrule(lr){6-8}
Dataset & Queries & BM25 ($a$) & WP-T ($b$) & WP-DE & BM25 ($a$) & WP-T ($b$) & WP-DE \\
\midrule
DBpedia & 400 & 0.318 & 0.400$^{a^+}$ & 0.392$^{a^+}$ & 0.373 & 0.416$^{a^+}$ & 0.408$^{a^+}$ \\
FiQA & 648 & 0.236 & 0.347$^{a^+}$ & 0.335$^{a^+}$ & 0.299 & 0.383$^{a^+}$ & 0.371$^{a^+}$ \\
NFCorpus & 308 & 0.338 & 0.365$^{a^+}$ & 0.351$^{b^-}$ & 0.287 & 0.300$^{a^+}$ & 0.292$^{a^+}$ \\
SciFact & 300 & 0.679 & 0.729$^{a^+}$ & 0.714$^{b^-}$ & 0.707 & 0.746$^{a^+}$ & 0.734$^{a^+,b^-}$ \\
Touche-2020 & 49 & 0.442 & 0.401$^{a^-}$ & 0.387$^{a^-,b^-}$ & 0.520 & 0.501$^{a^-}$ & 0.492$^{a^-,b^-}$ \\
TREC-COVID & 50 & 0.595 & 0.750$^{a^+}$ & 0.742$^{a^+}$ & 0.438 & 0.471$^{a^+}$ & 0.466$^{a^+}$ \\
\midrule

Avg. & -- & 0.435 & 0.499 & 0.487 & 0.437 & 0.469 & 0.461 \\
\bottomrule
\end{tabular}

\smallskip
\begin{tabular}{lrrr}
\toprule
\multicolumn{4}{c}{Mean Complete-Ranking Cost per Query} \\
\midrule
Method & Comparisons & Tokens (K) & Ranking Time (s) \\
\midrule
WP-T & 93.2 & 1,512.2 & 190.6 \\
WP-DE & 47.6 & 765.0 & 93.8 \\
\bottomrule
\end{tabular}
\vspace{-10pt}
\end{table}

Table~\ref{tab:beir-generalization} evaluates across six BEIR datasets. Averaged over all nine backbones, WP-DE improves over BM25 (Touche-2020 is an exception). As WP-T and WP-DE rerank identical BM25 candidate pools, their effectiveness differences arise from ranking order alone. WP-DE trails WP-T overall. The size of this gap varies across datasets: the difference is small on DBpedia, FiQA, NFCorpus, and TREC-COVID, but larger on SciFact and Touche-2020. However, WP-DE consistently and substantially reduces computation: lowering mean comparisons from 93.2 to 47.6, token consumption by 49.4\%, and ranking time from 190.6 to 93.8 seconds per query. These results show that WP-DE’s computational savings transfer across domains, while the associated effectiveness loss is dataset-dependent.

\subsection{Selection and Comparison-Scope Analysis}
\label{sec:selection-scope}

We next investigate whether WP-DE's bottom (worst candidate) updates might explain its top-10 effectiveness gap with WP-T, whether larger comparison windows can reduce computational cost, and how Whole-Pool trade-off changes across candidate pool sizes.

To examine the impact of bottom updates, in addition to WP-T and WP-DE, we evaluate a WP-DE variant that selects both the top and bottom candidates but updates only the top position, returning the bottom candidate to the pool. Relative to WP-DE, omitting the bottom update changes nDCG@10 by $+0.0004$, $-0.0034$, and $+0.0101$ for Q9B, L8B, and M8B, respectively. The gap to WP-T remains for Q9B and L8B, whereas both WP-DE variants outperform WP-T on M8B. These mixed effects show that removing bottom updates does not consistently recover top-10 effectiveness, suggesting that the bottom updates alone do not fully explain the differences.

\input{figures/window_sweep_three_panel}

Setwise methods use a window $w$ for comparison; where larger $w$ might reduce computational cost. We vary $w\in \{3,10,20,50,100\}$ for SW-TB with $N=100$ and $k=10$ to validate this assumption. Figure~\ref{fig:setwise-window-sweep} shows that larger windows indeed reduce the local comparisons required to select each top candidate. Increasing $w$ from 3 to 20 reduces mean comparisons from 451.6 to 48.6, while mean nDCG@10 changes only from 0.690 to 0.689. At $w=50$ and $w=100$, effectiveness dives, consistent with the greater difficulty of selecting one winner from more candidates. The $w=50$ setting also uses fewer tokens and less ranking time than $w=100$, showing that processing more candidates per call can offset the benefit of fewer comparisons.

At $w=100$, each comparison selects one top candidate from the entire remaining candidate pool, following WP-T's top-selection pattern with different labelling and parsing configurations. This connects wider comparison scope to Whole-Pool selection, where WP-DE adds a separate saving by selecting both a top and a bottom candidate per comparison, reducing the total comparisons needed to complete the ranking.

We further test how the Whole-Pool trade-off changes across candidate pool sizes. To this end, we vary the candidate-pool size $N$ from 10 to 100 for complete ranking. At every tested size, WP-DE remains consistently closer to WP-T than WP-B in both nDCG@10 and nDCG@N, while requiring fewer comparisons than either single-end variant. Its nDCG@N deficit relative to WP-T remains between 0.004 and 0.012, while comparison savings increase from 37.5\% at N=10 to 49.0\% at N=100. Thus, the effectiveness-efficiency trade-off persists as the candidate pool grows.

\subsection{Candidate Ordering, Labels, and Rank Fusions}
\label{sec:presentation-sensitivity}

We examine how WP-DE responds to changes in candidate ordering and candidate labels, and whether combining rankings from different orders improves effectiveness. These experiments use DL19 at $N=100$ with Q9B, L8B and M8B.

Figure~\ref{fig:presentation_sensitivity} (a) compares three candidate-ordering settings. \emph{Forward} starts with BM25 original candidate ordering and subsequently follows WP-DE's standard endpoint swaps. \emph{Reverse} presents a reversed copy of the current remaining candidate pool before each comparison, the final ranking is updated only through the standard WP-DE's endpoint method, \emph{Shuffle} presents a randomly permuted copy of the current remaining candidate pool before each comparison, using a deterministic seed derived from the query ID and remaining pool size. Each setting uses 50 comparisons per query. Reverse and Shuffle yield lower observed nDCG@10 than Forward on all three backbones; the largest decrease is 0.131 for L8B under Reverse and is statistically significant. These results indicate that effectiveness is influenced by the initial ordering of the candidate pool.

\input{figures/presentation_sensitivity_two_panel}

LLM rankers need to generate candidate identifiers as their output; the format of these identifiers may influence effectiveness. First, we use sequential numeric labels for candidates in the pool we call \emph{Canonical} identifiers. Then we produce a \emph{Opaque} variant which uses numeric identifiers unrelated to candidate positions. Figure~\ref{fig:presentation_sensitivity} (b) shows the impact of this change is small and inconsistent effects: slight gains for Q9B and L8B, but a decrease for M8B. It therefore provides no consistent effectiveness gain over the Canonical configuration.

We then combine two complete WP-DE rankings via two different rank fusion approaches to test if we can further improve effectiveness, with one from a Canonical run and one from a Shuffle run. \emph{Borda} sums rank-based points $N-r_j(d)$, while reciprocal rank fusion (\emph{RRF})~\cite{cormack2009rrf} sums $1/(60+r_j(d))$, where $r_j(d)$ is the one-based rank of candidate $d$ in ranking $j\in\{1,2\}$. Both rules favor candidates ranked highly across the two runs, ties follow the Canonical ranking. Producing the two rankings requires 100 comparisons in total, versus 50 for one run, while combining them requires no additional LLM comparisons. Both rules improve observed nDCG@10 over Canonical on all three backbones, showing that WP-DE's computational savings can support an additional run when higher top-10 effectiveness warrants more computation.

These experiments use the same retrieved candidates to compare different candidate orders, label schemes, and ways of rank fusions to combine the rankings. Across five repeated DL19 runs at $N=100$ for WP-T, WP-B, and WP-DE on the three backbones, nDCG@10 and nDCG@50 have standard deviations of 0.0000 at the reported precision. Thus, repeated runs under identical settings produce consistent scores, although changing the candidate order can impact the effectiveness.

%% file: figures/calls_vs_ndcg100.tex
\begin{figure}[t]
\centering
\includegraphics[width=\textwidth]{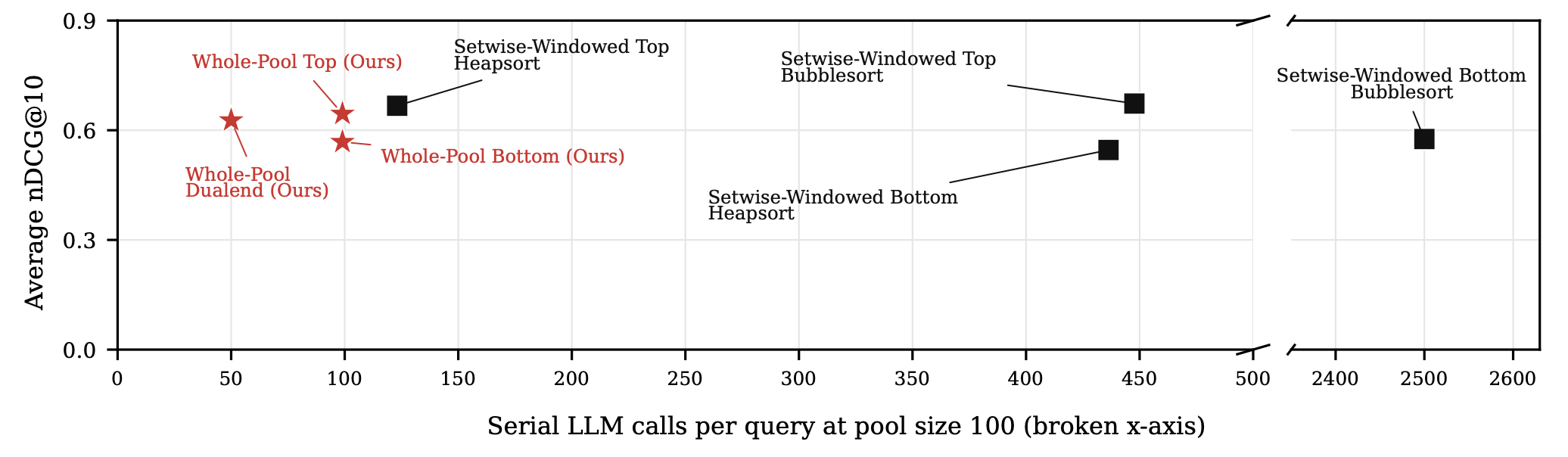}
\caption{Effectiveness-effeciency comparison (pool size 100), averaged across DL19, DL20 and nine backbones in Table~\ref{tab:dl-pool100-main}. WP methods construct complete rankings, three-passage SW methods target top-ten output. BM25 is omitted because it uses no LLM comparisons.}
\label{fig:calls-vs-ndcg10}
\vspace{-15pt}
\end{figure}

%% file: figures/window_sweep_three_panel.tex
\begin{figure}[t]
\centering
\includegraphics[width=\textwidth]{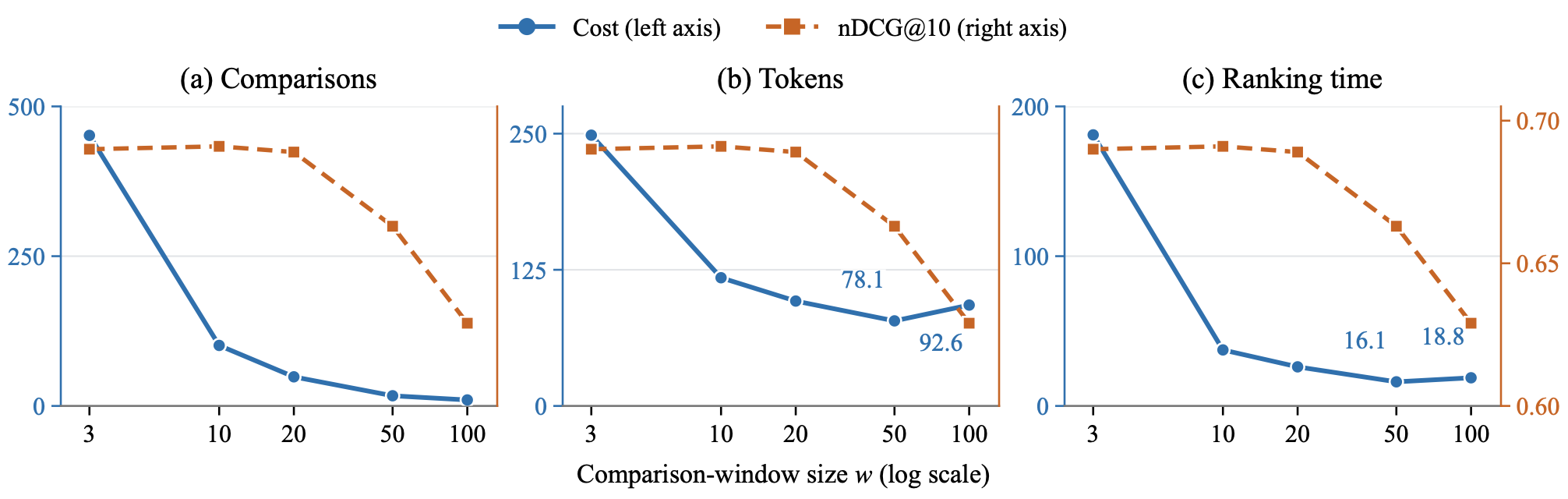}
\caption{Comparison-window sweep for SW-TB at fixed candidate-pool size $N=100$ and output depth $k=10$. Results average DL19/20 within each of Q9B, L8B and M8B, then over the three backbones. Left axes report (a) comparisons per query, (b) prompt and completion tokens per query in thousands, and (c) ranking time in seconds per query; right axes report nDCG@10.}
\label{fig:setwise-window-sweep}
\vspace{-10pt}
\end{figure}

%% file: figures/presentation_sensitivity_two_panel.tex
\begin{figure}[t]
\centering
\includegraphics[width=\textwidth]{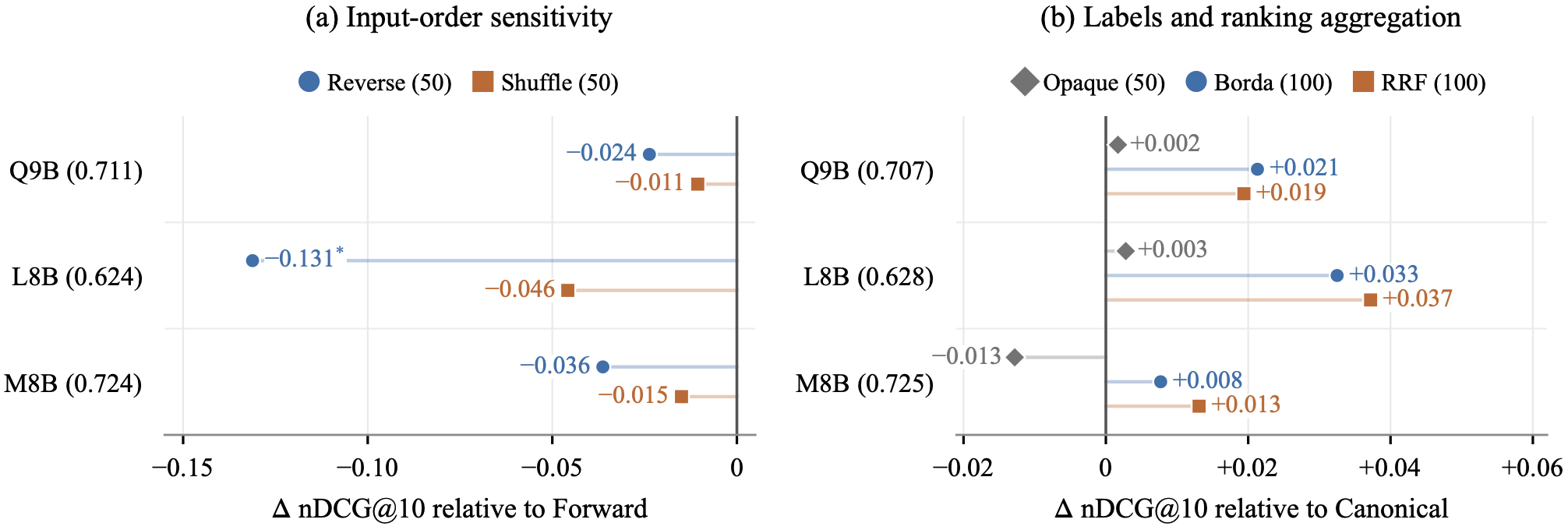}
\caption{Effects of candidate order, labels, and ranking combination on WP-DE effectiveness on DL19 at $N=100$. Changes in nDCG@10 are relative to (a) Forward for Reverse and Shuffle, and (b) a separately measured Canonical control for Opaque, Borda, and RRF. Parentheses beside backbone names give control scores; legend parentheses give comparisons per query, including both runs for Borda and RRF. The panels use different x-axis ranges. $^{*}$ denotes a significant difference from Forward under a two-sided paired t-test with Bonferroni correction ($\alpha=0.05$).}
\label{fig:presentation_sensitivity}
\vspace{-10pt}
\end{figure}

%% file: sections/conclusion.tex
\section{Conclusion and Limitations}
\label{sec:conclusion}

Our experiments demonstrate two complementary ways to make LLM-based reranking more efficient. Whole-Pool Setwise reduces the cost of repeated local sorting by selecting directly from all remaining candidates in the pool, while DualEnd further reduces the work required for complete ranking by resolving both the best and the worst candidate in each comparison. Compared with single-end Whole-Pool approaches, DuelEnd selection substantially reduces total number of LLM comparisons, token consumption, and ranking time while maintaining closely comparable complete ranking effectiveness on most tested backbones. These efficiency gains also generalize across the evaluated retrieval domains without additional ranking-specific training, although the associated effectiveness trade-off varies by backbone and dataset.

The benefits depend on the required ranking depth: DualEnd provides its largest comparison savings when a complete ranking is needed, whereas top-oriented alternatives can achieve stronger top-10 effectiveness. More broadly, computational cost is determined by both the number of comparisons and the amount of candidate text processed in each call, so fewer calls do not always translate into lower token consumption or latency. Ranking quality also depends on the order in which candidates are presented to the model, even when every remaining candidate is included in each comparison. Whole-Pool DualEnd selection therefore offers a practical efficiency–effectiveness trade-off for complete reranking, with scalability constrained by context capacity and the repeated cost of processing the remaining candidate pool.